\documentstyle[epsf,epsfig,mprocl]{article}

\newcommand {\be}{\begin{equation}}
\newcommand {\ee}{\end{equation}}

\newcommand {\cn}{{\cal N}}
\newcommand {\cC}{{\cal C}}
 
\begin{document}
 
\title{PERM: A MONTE CARLO STRATEGY FOR SIMULATING POLYMERS AND OTHER THINGS}

\author{P. GRASSBERGER$^{1,2}$, H. FRAUENKRON$^1$,
and W. NADLER$^1$ }
\address{ $^1$ HLRZ c/o Forschungszentrum J\"ulich, D-52425 J\"ulich, Germany\\
$^2$ Physics Department, University of Wuppertal, D-42097 Wuppertal, Germany }

\date{\today}

\maketitle
 
\abstracts{
We describe a general strategy, PERM ({\it P}runed-{\it E}nriched {\it R}osenbluth 
{\it M}ethod), for sampling configurations from a given Gibbs-Boltzmann distribution. The 
method is {\it not} based on the Metropolis concept of establishing a Markov process 
whose stationary state is the wanted distribution. Instead, it starts off building 
instances according to a biased distribution, but corrects for this by cloning ``good" 
and killing ``bad" configurations. In doing so, it uses the fact that nontrivial 
problems in statistical physics are high dimensional. Therefore, instances are built 
step by step, and the final ``success" of an instance can be guessed at an early 
stage. Using weighted samples, this is done so that the final distribution is strictly 
unbiased. In contrast to evolutionary algorithms, the cloning/killing is done without 
simultaneously keeping a large population in computer memory. We apply this in large scale 
simulations of homopolymers near the theta and unmixing critical points. In addition 
we sketch other applications, notably to polymers in confined geometries and to 
randomly branched polymers. For theta polymers we confirm the very strong logarithmic 
corrections found in previous work. For critical unmixing we essentially confirm the 
Flory-Huggins mean field theory and the logarithmic corrections to it computed by 
Duplantier. We suggest that the latter are responsible for some apparent violations 
of mean field behavior. This concerns in particular the exponent for the chain length
dependence of the critical density which is 1/2 in Flory-Huggins theory, but is claimed
to be $\approx 0.38$ in several experiments. }

\section{Introduction}

For many statistical physicists, ``Monte Carlo" is synonymous for the Metropolis 
strategy\cite{metrop} where one sets up an ergodic Markov process which has the desired 
Gibbs-Boltzmann distribution as its unique asymptotic state. There exist numerous refinements 
concerned with more efficient transitions in the Markov process (e.g. cluster flips\cite{swendsen} 
or pivot moves\cite{sokal}), or with distributions biased such that false minima are less
harmful and that autocorrelations are reduced (e.g. multicanonical\cite{berg} sampling and 
simulated tempering\cite{marinari}). But most of these schemes remain entirely within the 
framework of the Metropolis strategy.

On the other hand, there exists a very extensive literature on simulations of polymer systems
by completely different methods, starting already in the mid-50's and continuing until today. 
Polymers represent a challenge for simulations because of the constraints imposed by chain 
connectivity on the one hand, and steric hindrance on the other. These can lead to entanglement 
effects which seriously reduce the mobility of the monomers, and which can render Metropolis 
type schemes inefficient. 

The first such alternative method was invented by Rosenbluth and Rosenbluth.\cite{rosenbluth}
They observed that the exponential attrition in straightforward simulations of unbiased self 
avoiding random walks (SAW's) can be strongly reduced by simulating a biased sample, and 
correcting the bias by means of 
a weight associated to each configuration. The biased sample is simply obtained by replacing 
any ``illegal" step which would violate the self avoidance constraint by a random ``legal" one, 
provided such a legal step exists. More generally, assume we want to simulate a distribution 
in which each configuration $\cC$ is weighted by a (Boltzmann-)weight $Q(\cC)$, so that for 
any observable $A$ one has $\langle A\rangle = \sum_\cC A(\cC)Q(\cC)/\sum_\cC Q(\cC)$. If we 
sample unevenly with probability $p(\cC)$, then we must compensate this by giving a weight 
$\propto Q(\cC)/p(\cC)$, 
\be
   \langle A\rangle = \lim_{M\to \infty} {\sum_{i=1}^M A(\cC_i)Q(\cC_i)/ p(\cC_i)
                             \over \sum_{i=1}^M Q(\cC_i)/p(\cC_i) } \equiv 
                \lim_{M\to \infty} {1\over M} \sum_{i=1}^M A(\cC_i) W(\cC_i)\;.  
                                        \label{eq-rosen}
\ee

We call this the generalized Rosenbluth method. If $p(\cC)$ were chosen close to $Q(\cC)$, 
this would lead to importance sampling and obviously would be very efficient. But in general 
this is not possible, and Eq.(\ref{eq-rosen}) suffers from the problem that the sum is dominated by 
very few events with high weight. 

Consider now a lattice chain of length $N+1$ with self avoidance and with nearest neighbor 
interaction $-\epsilon$ between unbonded neighbors.
In the original Rosenbluth method, $p(\cC)$ is then a product,
\be
   p(\cC) \propto \prod_{n=1}^N {1\over m_n},              \label{eq2}
\ee
where $m_n$ is the number of free neighbors in the $n$-th step, i.e. the number of possible lattice 
sites where to place the $n$-th monomer (monomers are labeled $n=0,1,\ldots N$). Similarly, 
$Q(\cC)$ is a product, 
\be
   Q(\cC) = \prod_{n=1}^N e^{-\beta E_n},                  \label{eq3}
\ee
where $\beta = 1/kT$ and $E_n=-\epsilon\sum_{k=0}^{n-1}\Delta_{kn}$ is the energy of the $n$-th 
monomer in the field of all previous ones ($\Delta_{kn}=1$ if and only if monomers $k$ and $n$ 
are neighbors and non-bonded, otherwise $\Delta_{kn}=0$).

Obviously, $p(\cC)$ favors compact configurations where monomers have only few free neighbors. 
This renders the Rosenbluth method unsuitable for long chains, except near the collapse 
(`theta') point where simulations with $N\leq 1000$ are feasible on the simple cubic 
lattice.\cite{brun} In general we should find ways to modify the sampling so that ``good" 
configurations are sampled more frequently, and ``bad" ones less. The key to this is the 
product structure of the weights 
\be
   W(\cC_i) = {Q(\cC_i)/p(\cC_i) \over M^{-1}\sum_{k=1}^M Q(\cC_k)/p(\cC_k)} = 
          {1\over \hat{Z}_N} \prod_{n=1}^N w_n(\cC_i)
\ee
implied by Eqs.~(\ref{eq2}) and (\ref{eq3}). Here, 
$\hat{Z}_N = M^{-1}\sum_{k=1}^M Q(\cC_k)/p(\cC_k)$ is an estimate of the partition sum. 
A similar product structure holds in practically all interesting cases.

We can thus watch how the weight builds up while the chain is constructed step by step. 
If the partial weight (from now on we drop the dependence on $\cC$)
\be 
   W_n = \hat{Z}_n^{-1}\;\prod_{j=1}^n w_j\;,\qquad 1<n<N\;,
\ee
gets too large (i.e. is above some threshold $W^+$), we replace the configuration by $k$ copies, 
each with weight $W_n/k$. One of these copies is continued to grow, all others are placed on a 
stack for later use. Following Ref.\cite{wall} we call this `enrichment'. The opposite action, 
when $W_n$ falls below another threshold $W^-$ (`pruning'), is done stochastically: With 
probability 1/2 the configuration is killed and replaced by the top of the stack, while its 
weight is doubled in the other half of cases. 

In this way PERM (pruned-enriched Rosenbluth method \cite{perm1}) gives a sample with exactly 
the right statistical weights, independently of the thresholds $W^{\pm}$, the selection 
probability $p(\cC)$, and the clone multiplicity $k$. But its efficiency depends strongly on 
good choices for these parameters. Notice that one has complete freedom in choosing them, 
and can even change them during a run. Fortunately, reasonably good choices are easy 
to find (more sophisticated choices needed at very low temperatures are discussed in 
Ref.\cite{fold,stiff}). The guiding principle for $p(\cC)$ is that it should lead as closely 
as possible to the correct final distribution, so that pruning and enrichment are kept to 
a minimum. This is also part of the guiding principles for $W^{\pm}$. In addition, $W^+$ 
and $W^-$ have to be chosen such that roughly the same effort is spent on simulating 
any part of the configuration. For polymers this means that the sample size should 
neither grow nor decrease with chain length $n$. This is easily done by adjusting 
$W^+$ and $W^-$ `on the fly', see Ref.\cite{perm1} where a pseudocode is given for the 
entire algorithm.

In selecting the good and killing the bad, PERM is similar to evolutionary and 
genetic algorithms,\cite{holland}
to population based growth algorithms for chain poly\-mers,\cite{garel,higgs,velikson,orland}
to diffusion type quantum Monte Carlo algorithms,\cite{umrigar} and to the `go with the 
winners' strategy of Ref.\cite{aldous}. The main difference with the first three groups
of methods is that we do not keep the entire population of instances simultaneously 
in computer memory. Indeed, even on the stack we do not keep copies of good configurations 
but only the steps involved in constructing the configurations and flags telling us when to 
make a copy.\cite{perm1} In genetic algorithms, keeping the entire population in memory is needed 
for cross-overs, and it allows a one-to-one competition between instances. But in our case 
this is not needed since every instance can be compared to the average behavior of all others.
The same would be true for diffusion type quantum Monte Carlo simulations. The main 
advantage of our strategy is that it reduces enormously computer memory. This, together 
with the surprisingly easy determination of the thresholds $W^{\pm}$, could make PERM 
also a very useful strategy for quantum Monte Carlo simulations.

In section 2 we will apply PERM to single homopolymers on the simple cubic lattice
near their theta collapse temperature. This will be extended to semi-dilute solutions 
in Sec.~3, where we shall discuss critical unmixing of very long polymer chains. 
Finally, we shall sketch some further applications and improvements in Sec.~4. This 
includes what we call `markovian anticipation' and applications to $2-d$ SAW's,
to polymers in confined geometries, and to branched polymers.

\section{The $\Theta$ Collapse of Single Chains}

The $\Theta$ collapse is a tricritical phenomenon, described by the $n\to 0$ limit 
of the tricritical Landau-Ginzburg $O(n)$ model.\cite{degennes} Its upper 
critical dimension is, as for all tricritical $O(n)$ models, $d=3$. Therefore we must 
expect mean-field critical exponents in $d=3$, modified only by logarithmic corrections.
These corrections have been studied in great detail, partly due to some long-lasting 
controversies. The nowadays accepted results by Duplantier\cite{duplantier} include 
among others the following predictions: 

(i) The average end-to-end distance for a chain of $N$ elements grows at $T=T_\theta$ as 
\be
   R_N^2/N \approx const \left(1-{37\over363 \ln N}\right)          \label{r2-dupl}
\ee

(ii) The specific heat per monomer scales, again exactly at $T=T_\theta$, as 
\be
   c(T=T_\theta) = {1\over N} \left({1\over k_BT_\theta}\right)^2 
                 (\langle m^2\rangle-\langle m\rangle^2) 
                 \sim (\ln N)^{3/11}   \label{c-dupl}
\ee
where $m$ is the number of non-bonded nearest neighbor contacts.

(iii) In the collapsed phase $T<T_\theta$, the chain forms a globule with a bulk 
monomer density $\phi$ which scales as 
\be
       \phi \sim (T_\Theta-T) [-\log(T_\Theta-T)]^{7/11} .       \label{dupl-phi}
\ee

Here we shall only discuss chains modeled as SAW's on the 
simple cubic lattice with nearest neighbor attraction. For simplicity we choose this 
energy as $\epsilon = -1$, so that each non-bonded next neighbor contact contributes a 
factor $q=e^\beta$. 

Two-dimensional chains are treated with similar algorithms in Ref.\cite{theta2,spiral}, 
chains on other 3-d lattices and off-lattice chains are discussed in Ref.\cite{perm1,gh-off}. 

The simulations presented below show most dramatically the power of PERM. This depends 
mainly on the fact that the above logarithmic corrections are weak, and thus a simple 
random walk needs very little pruning and enrichment. To a very good approximation, 
the chain length $n$ performs during the simulation a random walk. Each addition of a 
monomer is a forward step of this walk, each pruning event is a jump backward to the 
next value of $n$ on the stack. The efficiency of the algorithm is directly measured 
by the diffusion constant $D$ of this walk, defined by
\be
   \langle (n_2-n_1)^2\rangle = 2D t,               \label{DD}
\ee
where $t$ is the number of steps (=subroutine function calls in the recursive 
implementation of Ref.\cite{perm1}) performed between lengths $n_1$ and $n_2$. 
The larger $D$, the faster a long chain is built up and disassembled again.
Measurements gave $D\approx 2000$ right at $T=T_\theta$. This is to be compared 
to $D = O(1)$ for other chain growth algorithms.\cite{reynolds-redner,beretti}

Results of simulations with $N=10,000$ are shown in Figs.~1 and 2. Both demonstrate 
clearly that the dominant behavior is mean field like, but the detailed predictions 
of Eqs.(\ref{r2-dupl}) and (\ref{c-dupl}) are not verified. Indeed, the corrections 
are much larger than predicted. It is not clear what is the source of these problems. 
It could be that higher order logarithmic corrections are important. But it could 
also be that multi-body forces between more than 3 monomers are important. 
Asymptotically, for $N\to\infty$, one expects three-body forces to be the only relevant 
ones (and this underlies Eqs.(\ref{r2-dupl}) and (\ref{c-dupl})), but at finite $N$ 
this might not be true. In any case, any 
 
  \begin{center}
    \vglue 1truemm
    \epsfig{file=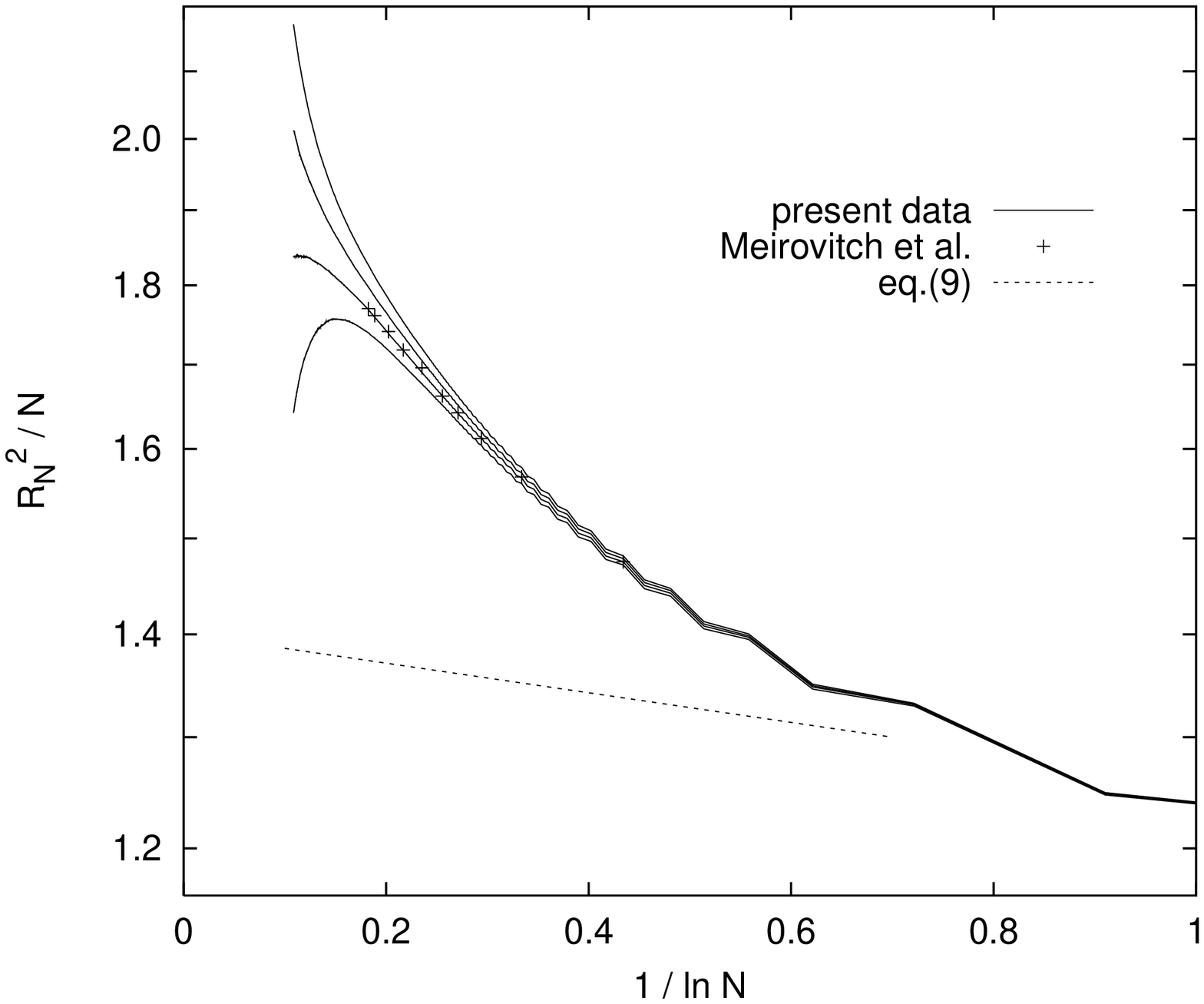, angle=270, width=7.8truecm}
  \end{center}
    {\small Figure~1: End-to-end swelling factor $R_N^2 /N$
      against $1/\ln N$, compared to previous\cite{meiro} MC estimates ($\Diamond$) 
      and to leading logarithmic prediction (dotted line).
      Curves are for $e^{1/kT}=$ 1.300, 1.305, 1.310, 1.315 (top to bottom). }

  \begin{center}
    \vglue 3truemm
    \epsfig{file=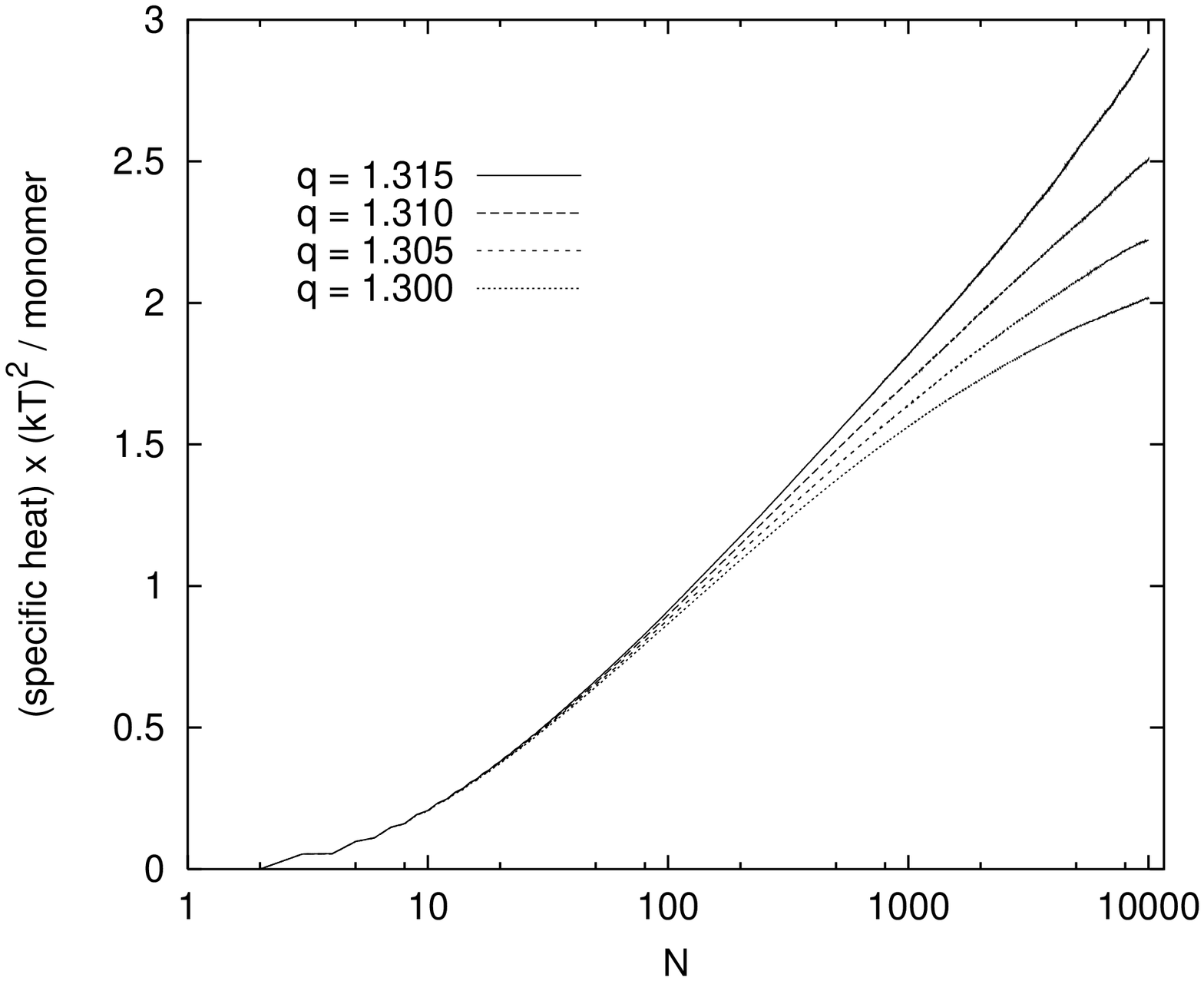, angle=270, width=7.8truecm}
  \end{center}
    {\small Figure~2: Variance of contact energies ($\sim$ specific heat)
      per monomer, for the same temperatures as in Fig.~1 ($q=e^{1/kT}$). Theory predicts
      $C/N \sim (\log N)^{3/11}$. }
  \vglue 3truemm

\begin{center}
  \epsfig{file=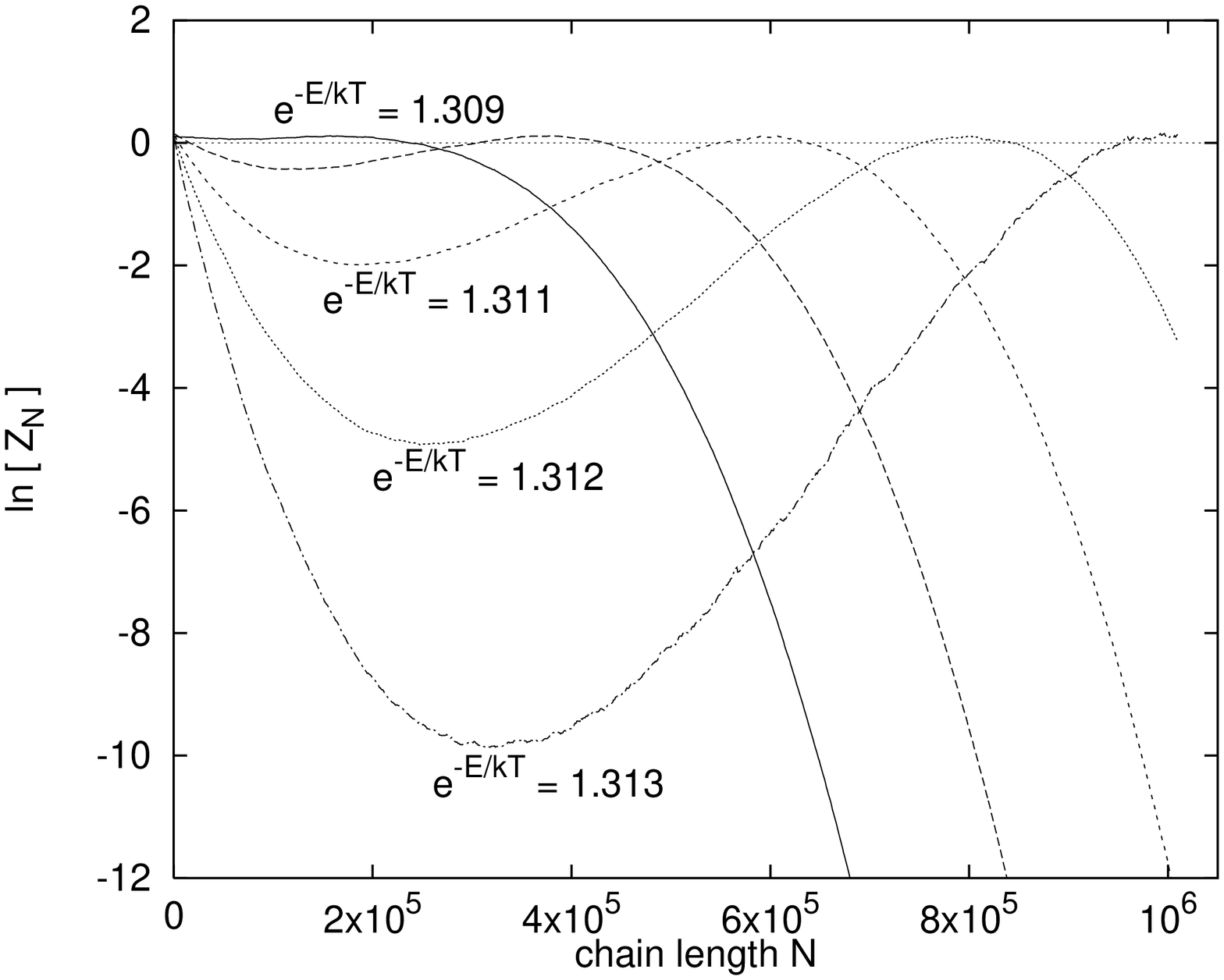, angle=270, width=7.9truecm}
\end{center}
  {\small Figure~3: Logarithms of partition sums (negative free energies) of single chains in a 
    box of size $256\times 256\times 256$. Notice that these are total free energies, 
    not free energies per monomer. Fugacities are adjusted such that 
    both maxima have the same height. The locations of the right-hand maxima give then
    estimates of $\Phi_\infty(T)$. \\
    Results of this procedure are shown in Fig.~4. }

\begin{center}
  \vglue 2truemm
  \epsfig{file=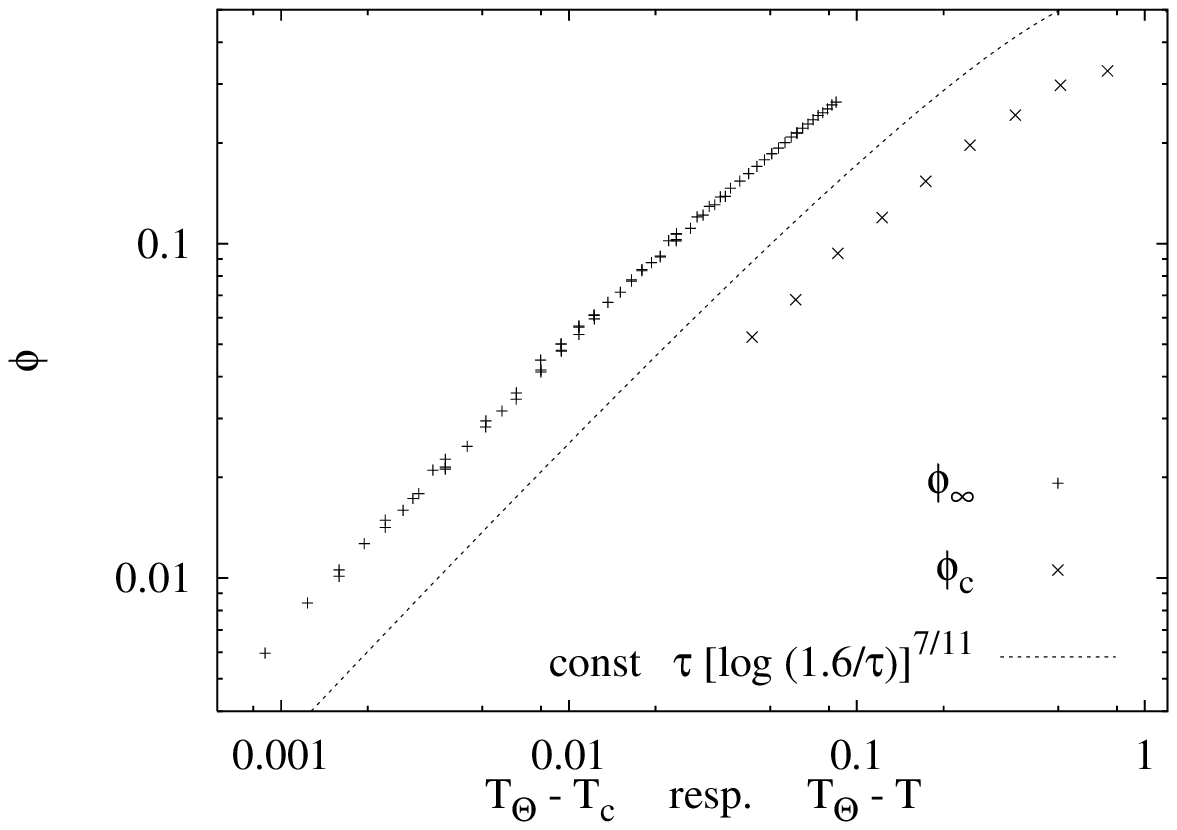, width=8.7truecm}
\end{center}
  {\small Figure~4: Plus signs (+): log-log plot of the infinite-$N$ monomer density $\phi$
    against $T_\theta - T$.\\
    Crosses ($\times$): log-log plot of critical monomer densities at finite $N$, plotted
    against $T_\theta - T_c(N)$, where $T_c(N)$ is the $N-$dependent critical temperature
    for unmixing.\\
    Dashed line: theoretical prediction, up to arbitrary factor. }
  \vglue 3truemm

\noindent
attempt\cite{booth} to interpret present 
experimental data by means of Eqs.(\ref{r2-dupl}) and (\ref{c-dupl}) is 
likely to give wrong conclusions.

In order to test prediction (iii) directly, one would have to simulate extremely long 
chains ($N\gg 10^6$), since otherwise results are strongly influenced by the surface of 
the globule. As an alternative, one can simulate systems without any surface by 
squeezing a chain into a cube with periodic boundary conditions. Above the $\Theta$-point, 
the free energy will be a (cap-)convex function of $N$, while it should be non-convex 
for $T<T_\theta$. From this it should be possible to estimate both $T_\theta$ and 
the density $\phi$. We were able to simulate chains of length up to $1.6\times 10^6$, 
in lattices of up to $2^{25}$ sites. Some typical results are shown in Fig.~3. They 
show that $T_\theta$ can indeed be estimated reliably, with results agreeing 
perfectly with those from Figs.~1 and 2. 
Values of $\phi$ obtained from Fig.~3 and similar plots are shown in Fig.~4, together 
with critical densities of multichain systems discussed in the next section. The 
dotted line in Fig.~4 represents the 
prediction (\ref{dupl-phi}). We see very good agreement, much better than expected 
after having seen Figs.~1 and 2.

\section{Critical Unmixing}

Far below the $\Theta$ collapse temperature, it is intuitively obvious that not only 
different parts of one long chain will coalesce, but also different chains will lump 
together and precipitate. For finite but large $N$ one expects therefore an unmixing 
transition at a critical temperature $T_c(N)$ which approaches $T_\theta$ from below 
when $N\to\infty$. For any fixed $N$ this will be a critical point in the Ising 
universality class, but additional scaling laws are expected as $N\to\infty$. A 
schematic phase diagram is shown in Fig.~5. 

\begin{center}
  \vglue 3truemm
  \epsfig{file=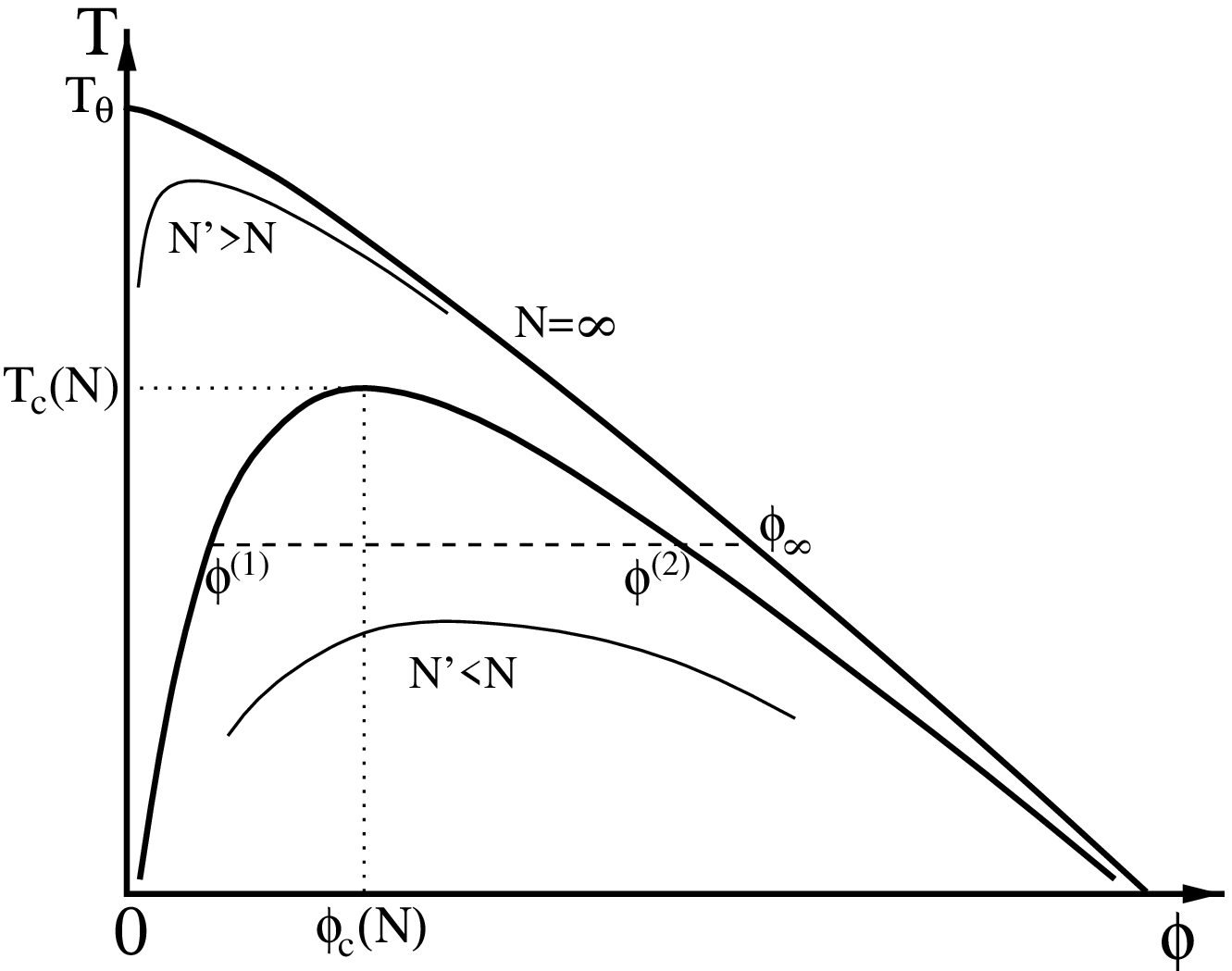, width=8.6truecm}
\end{center}
  {\small Figure~5: Schematic phase diagram for semi-dilute solutions of
    chain polymers. Uppermost curve: monomer concentration
    inside an infinitely large collapsed globule under zero outside
    pressure. Lower curves: coexistence curves for fixed chain
    length $N$. Curves are strictly monotonically ordered,
    with the coexistence curve for $N$ below that for $N'$ if $N<N'$.}
  \vglue 3truemm

For large $N$ one expects critical densities $\phi_c(N)$, deviations of $T_c(N)$ 
from $T_\theta$, and critical amplitudes $\phi^{(2)}-\phi^{(1)}$ to scale with $N$, 
\be
  \phi^{(2)} - \phi^{(1)} \sim (T_c-T)^\beta N^{-x_1}\;,  \label{dphi}
\ee
\be
  \phi_c \sim N^{-x_2}\;,                                 \label{phic}
\ee
\be
  T_\theta - T_c(N) \sim N^{-x_3}\;.                      \label{dtn}
\ee
Flory-Huggins theory\cite{degennes-book} predicts $x_1=1/4,\;x_2=x_3=1/2$. Notice 
that these predictions stand on rather different footings. Exponents $x_2$ and $x_3$ 
actually describe only properties which are not influenced by critical Ising-type 
fluctuations, and should therefore be correctly predicted by Flory-Huggins theory. 
In contrast, one has to expect that $x_1$ is heavily influenced by anomalous Ising 
scaling, since it appears in conjunction with $\beta$ which is different from its 
mean-field value in $d=3$. Indeed we had problems in measuring $x_1$ reliably.\cite{multi}
Therefore we shall discuss in the following only $x_2$ and $x_3$, and 
the chain swelling $R_N^2/N$. Naively one should expect that chains are ideal at 
the critical point.

Experiments\cite{perzynski,dobashi,shinozaki,chu,xia,izumi} suggest very strongly 
that $x_3$ is smaller than its mean-field value, $x_3 = 0.38\pm 0.01$. This problem 
has lead to numerous specula\-tions.\cite{sanchez1,sanchez2,muthu} In particular, it 
was suggested~\cite{sanchez1,muthu,sanchez2,chera} that chains are already partly 
collapsed at $T_c(N)$. For a review, see Ref.\cite{widom}.

Simulations with PERM are straightforward. We just have to place every $N$-th 
monomer not next to the previous one but at a random free site. In order to get the 
correct statistics (different chains are identical) we have to divide the partition 
sum by a factor $1/M!$ for a system with $M$ chains. We simulated chain lengths 
between $N=8$ and $N=4096$. The latter is much larger than in any other simulation 
(the longest being $N=1000$ in Ref.\cite{pana}). 

Except for $N=4096$, lattice sizes were 
such that the critical systems had roughly 50 to 100 chains. Although this is at least 
as large as in most previous simulations, it is not enough to be free from very large 
finite size corrections. In order to cope with them, we used a double histogram method 
as suggested in Ref.\cite{wilding-bruce} and applied to the present problem in 
Ref.\cite{wmb,pana}. The main advantage of this method is that it allows to use field 
mixing\cite{wilding-bruce} to get rid of the very strong asymmetry of the histogram 
and to reduce the problem essentially to the well understood Ising problem where the 
symmetry $m\to -m$ allows to pin down very precisely the critical point. But the 
efficiency of field mixing was checked in Ref.\cite{wmb,pana} only in so far as it 
rendered the distribution of chain numbers $M$ symmetric. The stronger requirement 
that also the joint $(M,E)$-distribution ($E$=energy) should be symmetric was not 
tested in Ref.\cite{wmb,pana} due to lack of sufficient statistics. In our 
simulations\cite{multi} we found that this symmetry is indeed badly violated. We have 
no explanation for this, and no remedy. It is mostly this problem which prevented us 
from obtaining reliable estimates of $x_1$, see the discussion in Ref.\cite{multi}. 
But other findings are much less affected by it.

Our first result is that chains are not collapsed at the critical point but slightly 
swollen. This agrees with Ref.\cite{wmb}. The swelling is only weakly dependent on the 
monomer concentration $\phi$. Results obtained exactly at the critical point are 
shown in Fig.~6. They show that the swelling agrees in the limit $N\to\infty$ perfectly 
with that of isolated chains.

Log-log plots of $\phi_c(N)$ versus $T_\theta-T_c(N)$, of $(T_\theta-T_c(N))/T_c(N)$ 
versus $N$, and of $\phi_c(N)$ versus $N$ are shown in Figs.~4, 7, and 8. From Fig.~7 
we see clearly that $x_3\approx 1/2$, with surprisingly small logarithmic corrections. 
This might be due to our choice of scaling variable. Using $(T_\theta-T_c(N))/T_\theta$ 
instead of $(T_\theta-T_c(N))/T_c(N)$, though being equivalent for large $N$, would give 
much worse scaling. Nevertheless, Fig.~7 suggests strongly that $x_3$ is equal to its 
mean field value. This is in marked contrast to $x_2$. 

\begin{center}
  \epsfig{file=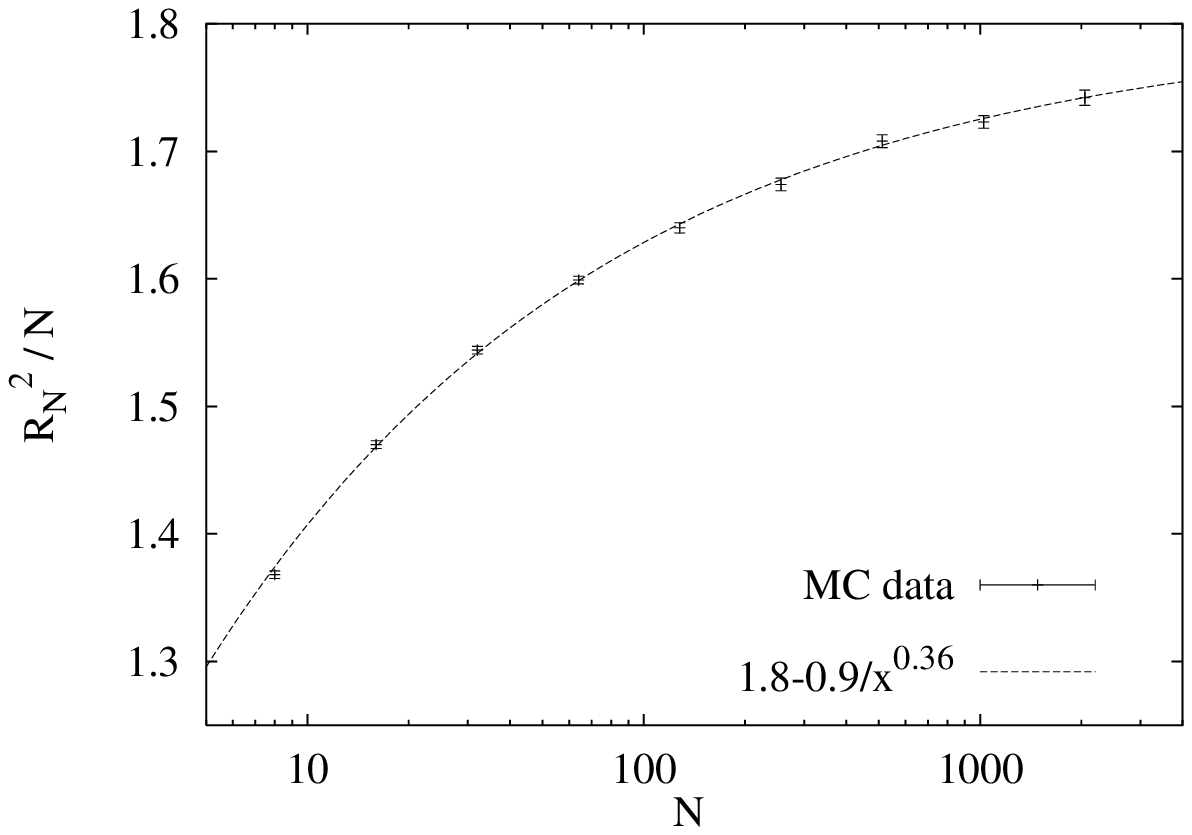, width=8.5truecm}
\end{center}
  {\small Figure~6: Swelling factors $R_N^2/N$ at the critical point
    plotted against $N$. The dashed line is a fit with the function
    $1.8-0.9x^{-0.36}$. This fit
    has no particular significance except for the fact that the limit
    $N\to\infty$ agrees with the swelling of infinitely long free $\Theta$
    polymers.}
  \vglue 2truemm

\begin{center}
  \vglue 2truemm
  \epsfig{file=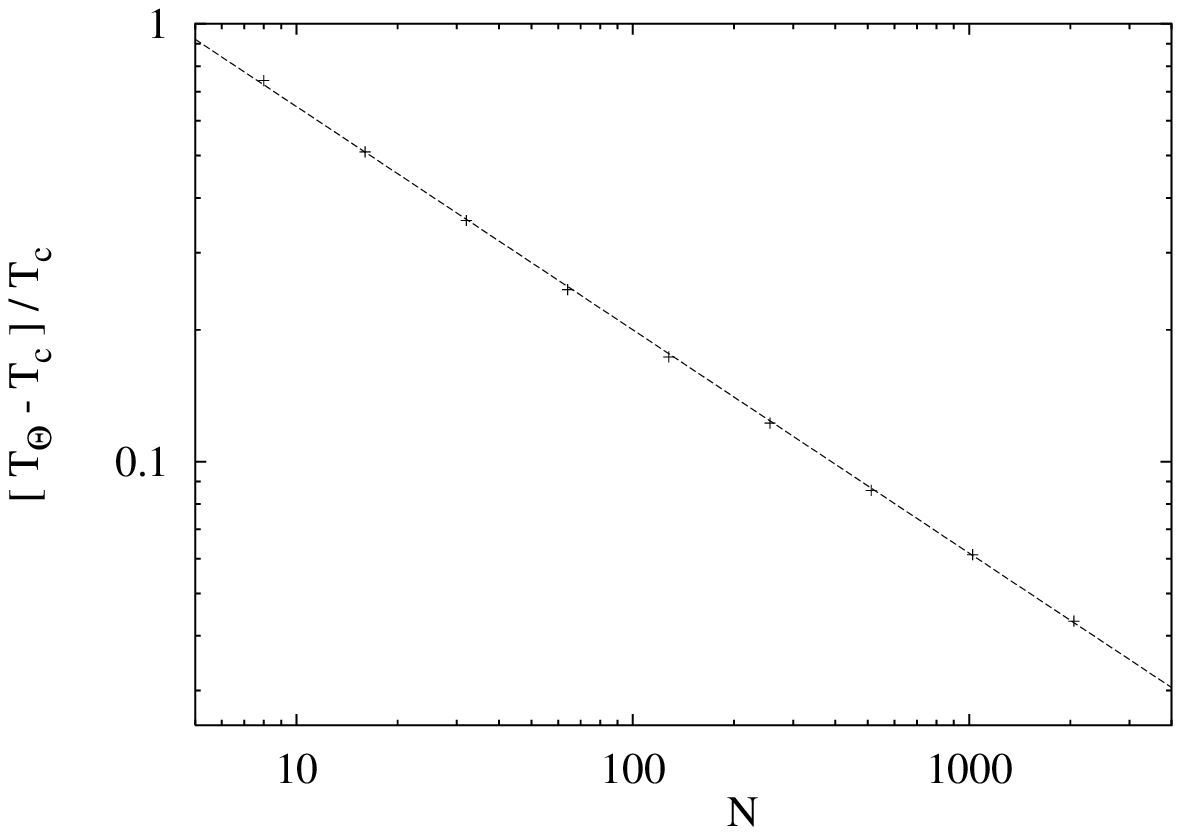, width=8.5truecm}
\end{center}
  {\small Figure~7: Log-log plot of $(T_\Theta - T_c(N))/T_c(N) $
    against $N$ (+). The dashed line has slope $-0.51$.}
  \vglue 3truemm

At first glance, Fig.~8 seems to give $x_2 = 0.385$, in perfect agreement with 
previous simulations\cite{wmb,pana} and with 
experiment.\cite{perzynski,dobashi,shinozaki,chu,xia,izumi} But we see already in 
Fig.~8 that there are strong deviations at small values of $N$, suggesting that this
is strongly affected by logarithmic corrections. This suspicion is confirmed by Fig.~4 
and by a strong theoretical argument~\cite{multi}: if $x_3=1/2$ (as confirmed by our 
simulations and by previous works\cite{wmb,pana}), and if coexistence curves are 
monotonically ordered as indicated in Fig.~5 (and as assumed by all previous authors), 
then $x_2$ is related to the dependence of $\phi(T)$ for isolated infinitely long 
chains: $\phi(T)\sim T^z$ with $z\leq 2x_2$. Thus $x_2<1/2$ would be inconsistent 
with the generally accepted value $z=1$ which we had also been confirmed by our 
simulations presented in Sec.~2. We thus conclude, in striking contrast to most
previous authors, that also $x_2$ has its mean field value, and that all observed 
deviations are (expected!) logarithmic corrections. Further arguments in favor of 
this interpretation, based on an explicit evaluation of the free energy, are given 
in Ref.\cite{multi}.

\begin{center}
  \vglue 2truemm
  \epsfig{file=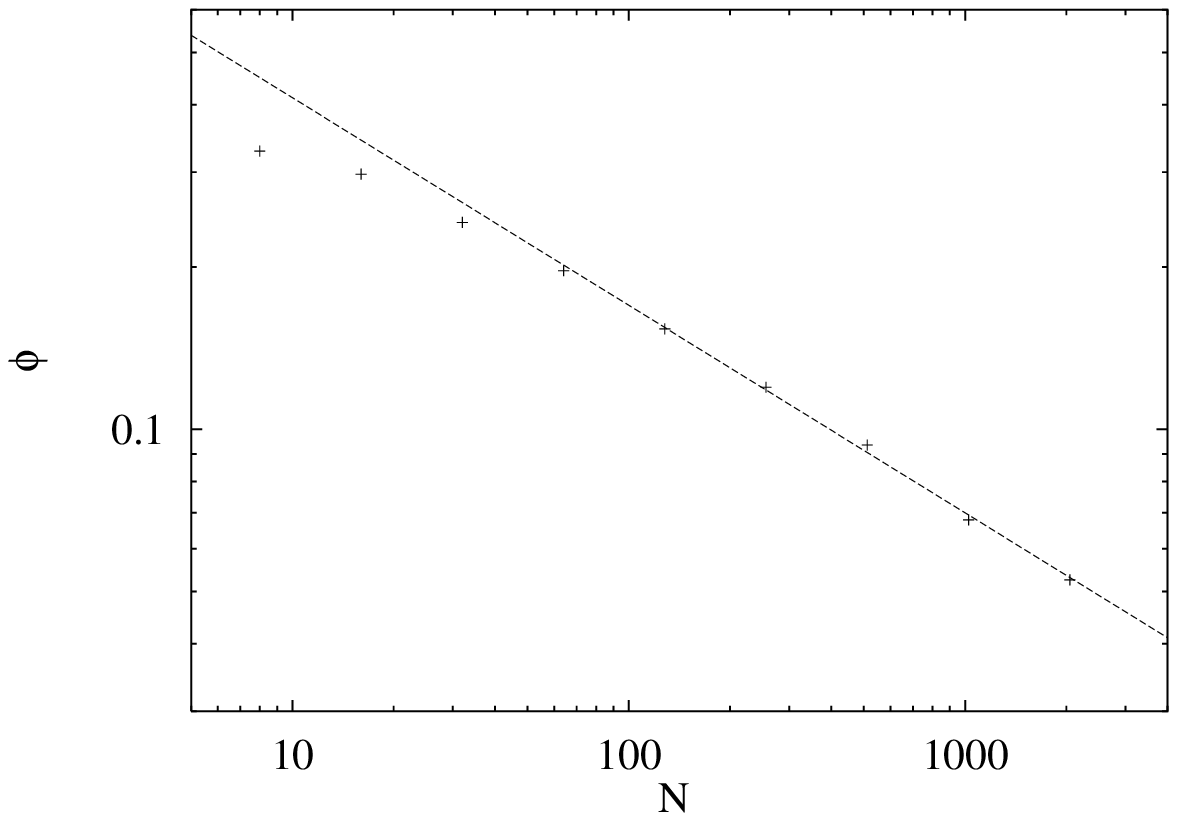, width=8.5truecm}
\end{center}
  {\small Figure~8: Log-log plot of $\phi_c(N)$ against $N$. The dashed
    line (slope $-0.385$) fits data for large $N$, but
    deviations at small $N$ are substantial.}
  \vglue 2truemm

\section{Further Developments} 

\subsection{Markovian Anticipation and 2-Dimensional SAW's}

As we had pointed out already in the introduction, PERM becomes most efficient if 
the a priori probabilities $p(\cC)$ are such that exactly the right distribution is 
obtained even without pruning and enrichment. In SAW's, this means that we should 
not choose randomly among the free neighbors when placing the next monomer. One way to 
``guide" the growth consists in looking ahead. In the `scanning method' of 
Meirovitch\cite{meiro} one looks $k$ steps ahead by checking all possible $k$-step 
extensions, and decides on the success of these extensions which single step to take next. 
This is efficient, but also very time consuming: the effort increases exponentially with $k$.

An alternative which costs hardly anything in terms of CPU time (but which requires 
rather large memory, if pushed to the extreme) consists essentially in looking back 
$k$ steps.  Assume we are dealing with a lattice with coordination number $\cn$.
During the initial steps of the simulation we build up two histograms $H_0(i)$ and 
$H_1(i)$ of size $\cn^{(k+1)}$ each. Here $i= 0,\ldots \cn^{(k+1)}-1$ labels all 
possible recent $k$-step histories and their one-step extensions. The first 
histogram $H_0(i)$ contains the weights of these $(k+1)$-step paths at the moment 
when the last step is made. The second histogram $H_1(i)$ contains the weights with 
which the same paths contribute much later (we used 100 steps in our simulations) 
to the partition sum. The ratio $H_1(i)/H_0(i)$ is therefore a direct estimate of 
how much the last step in path $i$ contributed to the final partition sum. Let us 
denote this last step as $j\; (=0,\ldots \cn-1)$ and the previous $k$ steps as $i_0$, 
so that $i=(i_0,j)$. Then we propose 
\be
   p(j|i_0) = { H_1(i_0,j)/H_0(i_0,j) \over \sum_{j'} H_1(i_0,j')/H_0(i_0,j')}
                                                 \label{ma}
\ee
as the best anticipation where to go next, based on a memory of $k$ steps.

Equation gives indeed a substantial improvement for 2-$d$ SAW's on all tested
lattices (square, triangular, honeycomb, Manhatten). Using $k$ such that 
$\cn^{(k+1)}\approx 10^6$ we get in all cases an increase by roughly a factor 10
in the effective diffusion constant $D$ (see Eq.(\ref{DD})). This means that the number 
of independent chains is increased by roughly one order of magnitude over the 
uniform choice.

Applications and details will be published elsewhere.\cite{markov} Here we just 
mention that this allows e.g. very efficient simulations in confined geometries. 
To illustrate this, we show in Fig.~9 a single chain of length $N=60,000$ in a 
slit of width 160. For convenience of plotting this was cut into 6 pieces. Such 
simulations allow significant tests of predictions about the monomer density 
in the vicinity of a hard wall.\cite{eisen}

\begin{center}
  \vglue 3truemm
  \epsfig{file=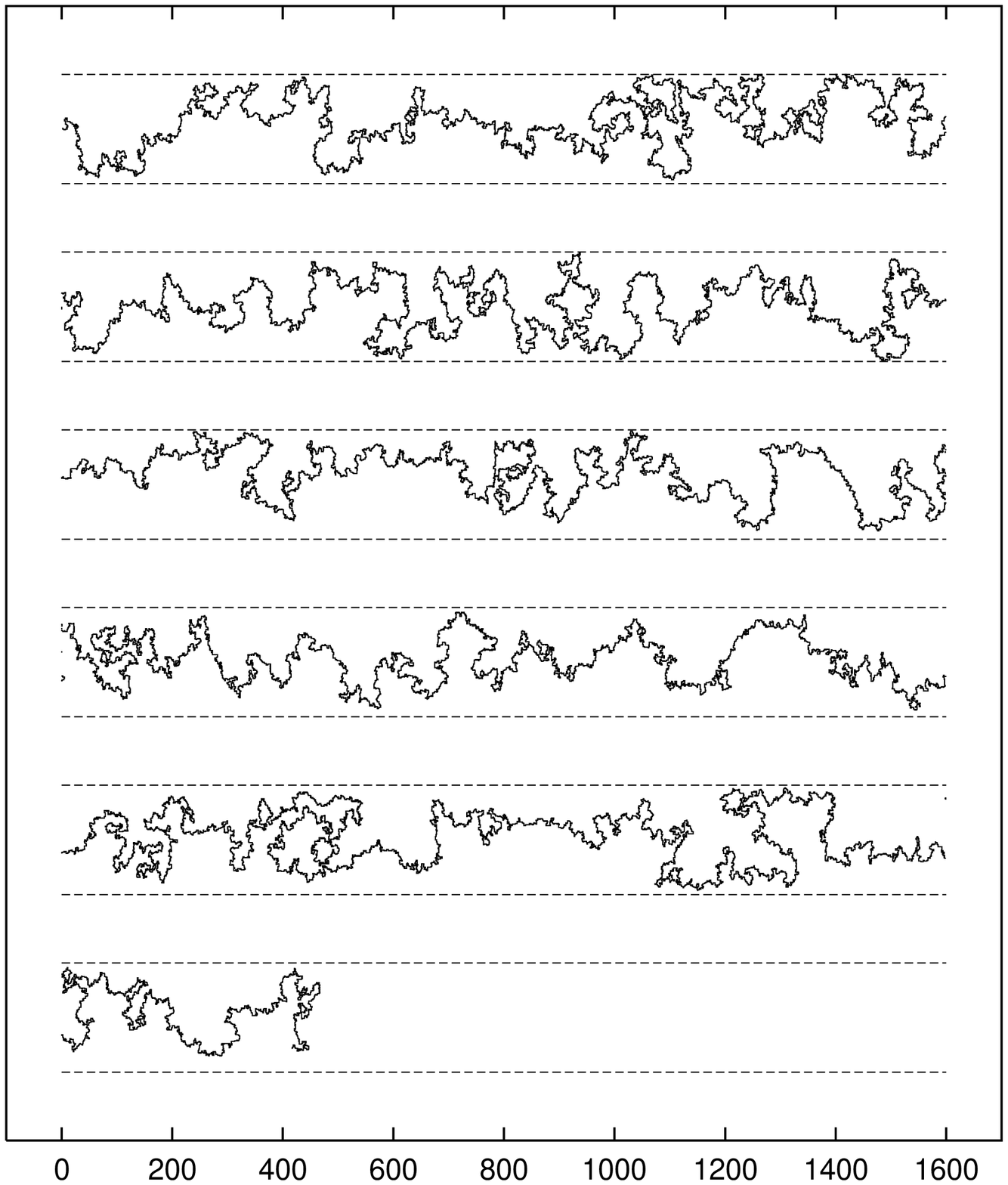, width=7.3truecm}
\end{center}
  {\small Figure~9: A single chain of length $N=60,000$ in a strip of width 160 of a 
    square lattice. In order to plot the configuration, it is cut in 6 pieces. }
  \vglue 3truemm

\subsection{Lattice Animals}

In our last example we want to illustrate that the strategy of PERM -- starting off 
with a biased selection, weighting it properly, and using these weights to redirect 
the selection by cloning and killing -- can also be implemented quite differently. 

Consider the set of all connected clusters $\cC_n$ of $n$ sites on a regular lattice, 
with the origin being one of these sites, and with a weight $Q(\cC_n)$ defined on 
each cluster. The ($n$-site) lattice animal problem is defined by giving the same 
weight to each cluster. The last requirement distinguishes animal statistics 
from statistics of percolation clusters. Take site percolation for definiteness, 
with `wetting' probability $p$. Then a cluster of $n$ sites with $b$ boundary sites 
carries a weight $p^n(1-p)^b$ in the percolation ensemble, while its weight in the 
animals ensemble is independent of $b$. In the limit $p\to 0$ this difference 
disappears obviously, and the two statistics coincide. Due to universality, we 
expect indeed that the scaling behavior is the same for any value of $p$ less than 
the critical percolation threshold $p_c$. 

While there exists no simple and efficient algorithm for simulating large animals 
(for a recent Rosenbluth type algorithm see Ref.\cite{care}), there exist very simple 
and efficient algorithms for percolation clusters. The best known is presumably the 
Leath algorithm\cite{leath} which constructs the cluster in a ``breadth first" 
(i.e., layer by layer) way, the easiest to implement is a recursive ``depth first" 
algorithm.\cite{sw}

Our PERM strategy consists now in starting off to generate subcritical percolation 
clusters by the Leath method, and in making clones of those growing clusters which 
contribute more than average to the animal ensemble.\cite{animals} Since we work 
at $p<p_c$, each cluster growth would stop sooner or later if there were no 
enrichment. Therefore we do not need 
explicit pruning. The threshold $W^+$ for cloning is chosen such that it depends 
both on the present animal weight and on the anticipated weight at the end of 
growth. 

\begin{center}
  \vglue 2truemm
  \epsfig{file=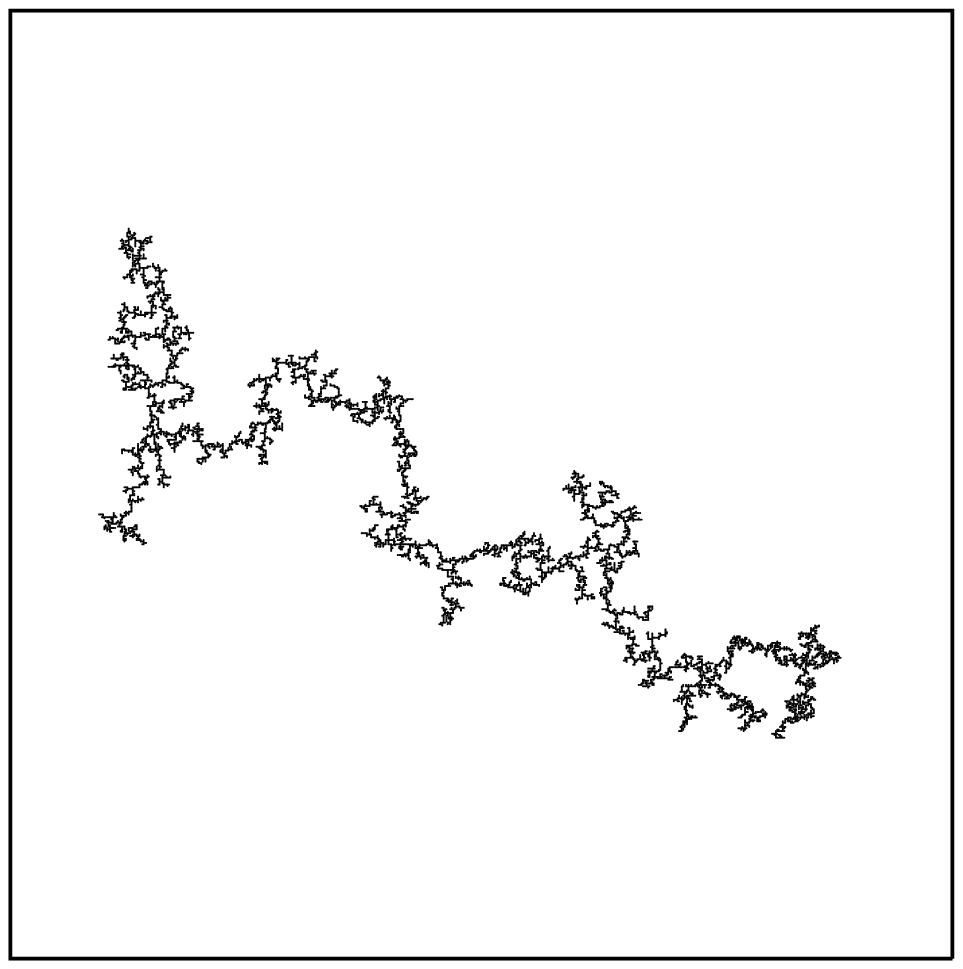, width=7.5truecm}
\end{center}
  {\small Fig.~10: A typical lattice animal with 8000 sites on the square lattice.}
  \vglue 2truemm

Usually, with growth algorithms like the Leath method, cluster statistics 
is updated only after clusters have stopped growing. But, as outlined below, one 
can also include contributions of still growing clusters. For percolation, this 
reduces slightly the statistical fluctuations of the cluster size distributions, 
but the improvement is small. On the other hand, this improved strategy is crucial 
when using PERM to estimate animal statistics.

Consider a growing cluster during Leath growth. It contains $n$ wetted sites, $b$ 
boundary sites which are already known to be non-wetted, and $g$ boundary sites 
at which the cluster can still grow since their status has not yet been decided 
(``growth sites"). This cluster will contribute to the percolation ensemble only 
if growth actually stops at all growth sites, i.e. with weight $(1-p)^g$. Since 
the relative weights of the percolation and animal ensembles differ by a factor 
$(1-p)^{(b+g)}$ (since now $b+g$ is the total number of boundary sites), this 
cluster has weight $W(\cC)\propto (1-p)^{-b}$ in the animal ensemble. If we would 
use only this weight as a guide for cloning, we would clone if $W(\cC)$ is larger 
than some $W^+$ which is independent of $b$ and $g$, and which depends on $n$
in such a way that the sample size becomes independent of $n$. 
But clusters with many growth sites will of course have a bigger chance to keep 
growing and will contribute more to the precious statistics of very large clusters. 
It is not a priori clear what is the optimal choice for $W^+$ in view of this, but 
numerically we obtained best results for $W^+ \propto (1-p)^g$. 

In this way we were able to obtain good statistics for animals of several thousand 
sites, independent of the dimension of the lattice. A typical 2-$d$ animal with 
8000 sites is shown in Fig.~10. We were also able to simulate animal collapse (when
each nearest neighbor pair contributes $-\epsilon$ to the energy), and animals 
near an adsorbing surface. Details will be published in Ref.\cite{animals}.

\section{Discussion}

We have presented a general strategy (PERM) for sampling from any given probability 
distribution. The main idea is to follow initially a biased distribution which 
is more easy to simulate. This bias is taken into account by re-weighting the 
sample, and these weights are used to interfere by pruning and cloning. For this 
to be possible it is needed that construction of instances is done in many 
small steps (which is always the case in statistical physics), and that the 
initial growth of weights is not misleading. It is mainly the second 
requirement which limits the usefulness of our approach. In the 3-$d$ Ising 
model, for instance, our approach gave rather mediocre results. But in other 
cases it is much more efficient. For polymers near the $\Theta$ collapse point, 
in particular, it seems by far the most efficient method presently known. 

The method shows some similarity to genetic and other evolutionary algorithms since 
good (`fit') instances are multiplied, while bad (`unfit') ones are killed. But 
in contrast to them this is done such that the wanted probability 
distribution is respected exactly, and it is done without the need for keeping 
large populations of instances in computer memory. We believe that it is mainly 
these features which make our strategy a promising candidate for further 
applications. Several such applications to toy protein models are discussed in these 
proceedings.\cite{fold,stiff,amphi} Further success will depend strongly on whether 
good non-trivial choices for $p(\cC)$ (such as in markovian anticipation, Sec.~3a) 
or for $W^{\pm}$ (such as in Sec.~3b) can be found in the specific problem at hand.
Another promising avenue to follow is the combination of our strategy with 
more conventional (Metropolis type) concepts. If this merge succeeds, PERM might 
lead to an efficient algorithms for finding native states of real proteins.

\bigskip
 
The authors are most grateful to Gerard Barkema for numerous 
discussions.

\end{document}